\documentclass[10pt,aps,pra,superscriptaddress,twocolumn,amsmath,amssymb,floatfix,
nofootinbib %this makes footnotes appear below the page, and not in the bibliography like the revtex class does
]{revtex4-1}

\usepackage[dvipsnames]{xcolor}
\usepackage{graphicx}
\usepackage[
colorlinks,
citecolor=blue,
linkcolor=blue,
urlcolor=blue,plainpages=false,pdfpagelabels
]{hyperref}

\begin{document}
\title{Casimir force between Weyl semimetals in a chiral medium}

\author{M. Bel\'en Farias}
\affiliation{Department of Physics
	 and Materials Science, University of Luxembourg, 1511 Luxembourg, Luxembourg}
\author{Alexander A. Zyuzin}
\affiliation{Department of Applied Physics, Aalto University, P. O. Box 15100 FI-00076 AALTO, Finland}
\affiliation{Ioffe Physical-Technical Institute, 194021 St. Petersburg, Russia}
\author{Thomas L. Schmidt}
\affiliation{Department of Physics
		and Materials Science, University of Luxembourg, 1511 Luxembourg, Luxembourg}

\date{\today}

\begin{abstract}
We study the Casimir effect in a system composed of two Weyl semimetals (WSMs) separated by a gap filled with a chiral medium. We calculate the
optical response of the material to chiral photons in order to calculate the Casimir force. We find that if the medium between the two WSMs is a Faraday material, a repulsive Casimir force can be obtained, and for realistic experimental parameters its magnitude can be greatly enhanced at short distances of a few microns Moreover, in the system under consideration various parameters can be modified. Some of them are intrinsic to the materials employed (the absolute value of the Hall conductivity of the WSM, the Verdet constant of the Faraday material), while some of them can be manipulated externally even for a fixed sample, such as the external magnetic field and the orientation of the plates which determines the sign of their conductivity. Suitable combinations of these parameters can be used to switch from attraction to repulsion, and to place
the trapping distance, in which no force acts on the plates, at any desired distance between them.
\end{abstract}

\maketitle

\section{Introduction}

The Casimir effect \cite{casimir1948}, i.e., the existence of a force between two neutral bodies due to the quantum fluctuations of the vacuum electromagnetic (EM) field, is one of several intuition defying consequences of quantum mechanics. The measurement of this macroscopic effect represented a major achievement of quantum field theory \cite{lamoreaux2004casimir}. Nearly seven decades after its initial prediction, there is still a high interest in the study of both the static Casimir effect \cite{milton2004casimir} and its out-of-equilibrium counterpart: the dynamical Casimir effect and quantum friction \cite{ milton2016reviewfriction, review_friction}. Lately, these phenomena have been studied not only from the field-theoretical viewpoint, but also from a materials perspective \cite{woods2016materials}. As new materials with unique properties were discovered, their implications for these forces have been studied \cite{rodriguez2011casimir,rodriguez2014repulsive,rodriguez2017casimir,farias2017quantum,farias2018quantum,marachevsky2019casimir}, with the aim of enhancing them, reducing them, or use them to probe intrinsic properties of the materials under consideration.

One of the primary objectives that has been pursued lately in the equilibrium case is to achieve a repulsive Casimir force \cite{fialkovsky2018quest}. This objective originates not only from a fundamental interest,
but also from the point of view of nanotechnology, where Casimir forces can have detrimental consequences:
the Casimir interaction usually dominates at very small distances and, being attractive in the majority of cases, system parts tend to stick together, which might cause micro- and nanoelectromechanical systems (MEMS and NEMS) to stop working.
The quest for Casimir repulsion dates back many years \cite{milton2012repulsive}, and it was studied both between two macroscopic plates as well as between a macroscopic plate and a microscopic particle (Casimir-Polder repulsion). Casimir repulsion was successfully detected experimentally \cite{munday2009measured} in a system with planar geometry in which the three materials fulfilled the relation $\epsilon_1 < \epsilon_2 < \epsilon_3$, where $\epsilon_{1,3}$ are the dielectric constants of the outer materials and $\epsilon_2$ the one of the medium in between. Further proposals have involved different metamaterials featuring uniaxial and biaxial magnetodielectric anisotropies \cite{rosa2008casimir}. More recently, efforts are shifting in the direction of topological insulators, materials with an insulating bulk but conducting surfaces. A thorough review of the systems considered and the methods used for the calculations can be found in Ref.~\cite{fialkovsky2018quest}.

The work presented in this paper is based on two recent proposals that showed promising results. In the first one, Casimir repulsion was predicted between two Weyl semimetals (WSL) \cite{wilson2015repulsive}. These materials fall within the broader category of topological nodal semimetals, a novel class of materials with topologically nontrivial electronic structure and a gapless bulk. Those are 3D materials whose crystalline structure results in crossings of electronic bands near the Fermi level, which give rise to Weyl cones and low-energy quasiparticles obeying a Weyl Hamiltonian \cite{armitage2018weyl}. They are the first physical realization of Weyl fermions, and exhibit many effects predicted by high-energy physics. The crossing points are protected against backscattering by impurities, so these systems remain gapless even under small perturbations. They also exhibit remarkable topological properties, for instance, surface states in the form of Fermi arcs \cite{wan2011topological}, which are impossible to realize in purely two-dimensional systems.

Topological nodal semimetals can be classified according to the positions of the Weyl cones in momentum space: Dirac semimetals, the first realized experimentally, contain two overlapping Weyl cones, whereas in WSMs they are separated, either in momentum space (in the case of time-reversal symmetry breaking) or in energy (for broken spatial inversion symmetry). This symmetry-breaking is what allows for the existence of Casimir repulsion, and is what was exploited by Wilson \textit{et al.}~\cite{wilson2015repulsive}, who considered a system of two identical time-reversal symmetry breaking WSMs. There is also a very recent work where these results have been extended to type-II WSMs, and include the effect of surface currents \cite{rodriguez2019signatures}.

A second promising proposal involves two perfectly conducting plates and a chiral medium filling the gap between them.
In a chiral medium, photons with different chiralities have different propagation velocities, thus breaking inversion symmetry. This allows for Casimir repulsion, as was pointed out by
Jiang \textit{et al.}~\cite{jiang2019chiral}, who in order to calculate such a repulsive force, have developed a non-reciprocal Green's function method to obtain Lifshitz's formula \cite{book_milonni} for the Casimir force. They found the Casimir energy between two plane plates in non-reciprocal media, separated by a distance $a$ along the $z$ direction, to be given by
	\begin{equation} \label{eq:Lif}
E_C= \hbar \int_0^\infty \frac{d\omega}{2 \pi} \int \frac{d^2 k_\parallel}{(2 \pi)^2} \ln \det \left( \mathbb{I} - R_B U_{BA} R_A U_{AB}\right) \, ,
\end{equation}
where $\omega$ is the imaginary frequency and $\textbf{k}_\parallel = (k_x,k_y)$ is the momentum in the $xy$ plane parallel to the surfaces of the plates. $R_A$ is the reflection matrix of the plate filling the space $z<0$, and $R_B$ the reflection matrix of the plate filling the space $z>a$, while $U_{AB} (U_{BA})$ represents the translation matrix from $B$ to $A$ ($A$ to $B$).

\begin{figure}[t]
	\includegraphics[width=\columnwidth]{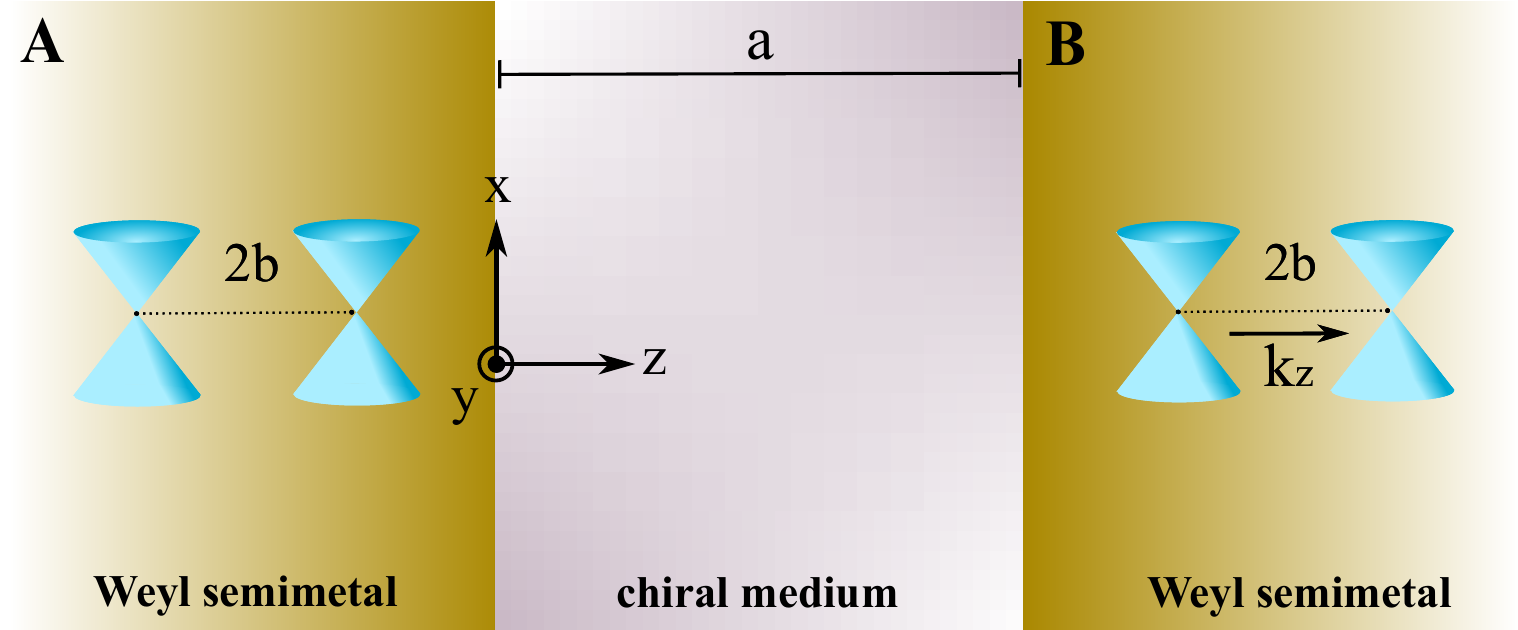}
	\caption{\label{fig:figure}Scheme of the system under consideration: two semi-infinite Weyl semimetals, with Weyl cones split in the $\hat{z}$ direction, are separated by a distance $a$. A medium with chiral properties fills the gap between them.}
\end{figure}

With these motivations, we will consider in this work a system composed of two WSMs separated by a gap filled with a chiral medium, as depicted in Fig.~\ref{fig:figure}. We will consider the Weyl cones
to be split in the same direction as the separation between plates, which results in a system with rotational invariance. This already hints at the fact that chirality has an important role to play in this type of WSM. In fact, in Ref.~\cite{fukushima2019anomalous} it was pointed out that WSMs can be thought as a particular type of chiral material. We will contemplate the different possibilities for the medium in between, and study the effects of the interplay between its properties and those of the WSMs on the Casimir force.

This paper is organized as follows: in Sec.~\ref{sec:wsmchiral}, we study the electrodynamics of WSMs in the presence of circularly polarized light. In Sec.~\ref{sec:chiralmedia}, we review the different types of chiral materials and their properties, to later combine these results in Sec.~\ref{sec:casimirforce} to calculate the Casimir force between two WSMs with a Faraday material filling the gap between then. Lastly, in Sec.~\ref{sec:conc} we present our conclusions.

\section{Weyl Semimetals and chiral light} \label{sec:wsmchiral}

\subsection{Electrodynamics inside the material}

The theory of electrodynamics inside a WSM has been developed in detail in Refs.~\cite{grushin2012consequences} and \cite{wilson2015repulsive}. The starting point is the action $S = S_0 + S_A$ for the electromagnetic field. The part $S_0$ is the ordinary Maxwell action
\begin{equation}
    S_0= - \frac{1}{4} \int d^3r\, dt\, F_{\mu \nu} F^{\mu \nu} - \frac{1}{c}\int d^3\, dt\,  A_\nu j^\nu\, ,
\end{equation}
where $j^\nu = (c\rho, \textbf{j})$ denotes a 4-vector ($\nu = 0,1,2,3$) consisting of the charge and current densities. The electromagnetic strength tensor is $F_{\mu \nu}= \partial_\mu A_\nu - \partial_\nu A_\mu$. The part $S_A$ is the axionic term \cite{zyuzin2012topological}
\begin{equation}
S_A=\frac{e^2}{32 \pi^2 \hbar c} \int d^3r \, dt \, \theta(\textbf{r},t) \, \epsilon^{\mu \nu \alpha \beta} F_{\mu \nu}F_{\alpha \beta} \, ,
\end{equation}
where $\epsilon^{\mu \nu \alpha \beta}$ is the fully anti-symmetric tensor and $\theta(\textbf{r},t)=2 \textbf{b} \cdot \textbf{r} - 2 b_0 t$. Here, $2 \textbf{b}$ is the separation between Weyl cones in momentum space and $2 b_0$ is their energy offset. By writing the Euler-Lagrange equations of motion for the vector potential $A_\mu$,
\begin{equation}
\label{eq:EL}
\partial_\nu F^{\nu \mu} + \frac{e^2}{8 \pi^2 \hbar c} b_\nu \epsilon^{\nu \mu \alpha \beta} F_{\alpha \beta} = \frac{1}{c}j^{\mu} \, ,
\end{equation}
one obtains modified inhomogeneous Maxwell equations, whereas the homogeneous ones are automatically satisfied due to the definition of $F_{\mu \nu}$.

From now on, we will consider the case where only time-reversal symmetry is broken, i.e., $b_0=0$. By preserving inversion symmetry in the WSM,
one obtains two Weyl cones with opposite chiralities at the same energy \cite{armitage2018weyl}. Specifically, we will assume $\textbf{b}=b \hat{z}$. This ensures that there are no surface Fermi arc states in the $z=0$ plane, so the Casimir force between plates separated along the $z$ direction will determined by the bulk states only. Casimir forces in other directions, in contrast, are affected by the surface states \cite{morali2019fermi}. Moreover, this choice of setup will preserve rotational symmetry in the $xy$ plane.

This means that the WSM under consideration is described by four modified Maxwell's equations \cite{grushin2012consequences}
\begin{align} \label{eq:Maxwell}
&\nabla \cdot \textbf{E} = \rho- \frac{e^2 b}{2 \pi^2 \hbar c} \alpha \, \hat{z} \cdot \textbf{B} \, \\
&\nabla \times \textbf{B} = \frac{1}{c} \frac{\partial \textbf{E}}{\partial t} +\frac{1}{c}\textbf{j}+ \frac{e^2 b}{2 \pi^2 \hbar c}  \hat{z} \times \textbf{E}  \, , \\
&\nabla \cdot \textbf{B}=0 \, , \\
&\nabla \times \textbf{E}= - \frac{1}{c} \frac{\partial \textbf{B}}{\partial t} \, ,
\end{align}
where we can identify the Hall conductivity as ${\sigma_{xy}=-\sigma_{yx}=e^2 b / 2 \pi^2 \hbar}$ and the correction to the charge density $- \frac{e^2 b}{2 \pi^2 \hbar c} \alpha \, \hat{z} \cdot \textbf{B}$. Hence, the axionic action generates a source term in Ampere's and Gauss's law. We also note that the ferromagnetic Weyl semimetal is a gyrotropic medium with the gyrotropy parameter being proportional to the separation $b$ of the Weyl cones in momentum space \cite{landau2004electrodynamics}.

We proceed to solve the modified Maxwell equation in Fourier space by using $\textbf{E}(\textbf{r}, t) = \textbf{E}(\textbf{k}, \omega) e^{i \textbf{k} \cdot \textbf{r}- i \omega t}$, and analogously for $\textbf{B}(\textbf{r}, t)$. An effective dielectric function $\epsilon (\omega) $ can be introduced to describe the macroscopic response of the material \cite{grushin2012consequences}
by defining the displacement field as $\textbf{D}(\textbf{k},\omega)=\epsilon (\omega) \textbf{E}(\textbf{k},\omega)$ with
\begin{equation}\label{eq:eps}
{\epsilon}(\omega)=\left(
\begin{array}{c c c}
\epsilon_0& i{\sigma_{xy}}/{\omega} & 0 \\
- i {\sigma_{xy}}/{\omega} & \epsilon_0 & 0 \\
0 & 0 & \epsilon_0
\end{array}
\right) \, .
\end{equation}
where at zero temperature and clean semimetal with two Weyl points one has \cite{lv2013dielectric}
\begin{align}
\epsilon_0 (\omega) = 1 + \frac{\alpha}{3\pi} \frac{c}{v_F}\left[\ln\left|\frac{\Lambda^2}{4\mu^2-\omega^2}\right|-\frac{4\mu^2}{\omega^2}+i\pi\Theta(\omega-2\mu) \right],
\end{align}
in which $\alpha = e^2/\hbar c$ is the fine structure constant.

In the previous equation, $v_F$ is the Fermi velocity for the excitations of the material, and we have introduced a cutoff $\Lambda$, whose maximum can be estimated as ${\Lambda \sim v_F b}$. When rotating to imaginary frequencies $\omega = i \xi$ as required by Lifthitz's equation and for vanishing chemical potential, we can write
\begin{equation}
\epsilon_0 (i \xi)= 1 + \frac{\alpha}{3\pi}  \frac{c}{v_F} \ln\left(\frac{\Lambda^2}{\xi^2}\right) \,.
\end{equation}
As will be shown in the following sections, the relevant frequencies contributing to the Casimir force are of the order $\xi \approx c/a$. This means that for $v_F/c \approx 10^3$, for distances between materials $a$ of the order of $1 \mu \text{m}$ and distances between Weyl cones such that $1/b \approx 1 \text{nm}$, the argument of the logarithm can be estimated as $(v_F/c)^2(a b)^2 \approx 1$. The prefactor in front of the logarithm will be as well of order one, so that we can approximate $\epsilon_0 \approx 1$ up to logarithmic corrections.
Our results, then, shall remain valid as long as we consider distances $a \sim c/\Lambda$.

With the definition~\eqref{eq:eps}, the two first equations in \eqref{eq:Maxwell} can be combined into a wave equation for the displacement field,
\begin{equation}
\label{eq:waveeq}
\left[ \textbf{k} \otimes \textbf{k} - k^2 \mathbb{I} \right] \epsilon^{-1} (\omega) \textbf{D}(\textbf{k},\omega)= - \frac{\omega^2}{c^2} \textbf{D}(\textbf{k},\omega) \, ,
\end{equation}
where $k^2 = k_x^2 + k_y^2 + k_z^2$.
We can obtain the dispersion relation and the allowed polarizations inside of the material as the solutions to the wave equation above.
For the system to have nontrivial solutions, the determinant of the matrix
\begin{equation}
\mathcal{M}(\omega) = \left[ \textbf{k} \otimes \textbf{k} - k^2 \mathbb{I} \right] {\epsilon}^{-1} (\omega) + \omega^2 \mathbb{I}
\end{equation}
must vanish. Here, we have taken $c=1$ and will do so in the rest of the paper unless otherwise stated.
The solutions to this equation provide the dispersion relation inside the WSM \cite{grushin2012consequences,wilson2015repulsive}:
\begin{eqnarray} \label{eq:disp}
\omega_\pm^2(\textbf{k}) = k^2 + \frac{ \sigma_{xy}^2}{2} \pm  \sqrt{{k_z^2 \sigma_{xy}^2} + \frac{\sigma_{xy}^4}{4}} \, .
\end{eqnarray}
Setting $\omega=\omega_\pm$ will ensure that Eq. \eqref{eq:waveeq} has non-trivial solutions,
which are the allowed polarizations inside the material.
The (unnormalized) polarization vectors corresponding to the energies $\omega_\pm$ are
\begin{equation}
\label{eq:polarizations}
\textbf{D}_\pm = \omega_\pm k_z \left( k^2  -  \omega_\pm^2 \right) \hat{e}_1 - i k  \sigma_{xy} (\omega_\pm^2 - k_x^2) \hat{e}_2
\end{equation}
where $\hat{e}_1 = \hat{y} \times \hat{k}$ and $\hat{e}_2 = \hat{y}$ are the transverse electric (TE) and transverse magnetic (TM) directions, respectively. Without loss of generality, we have chosen the $y$ axis such that $k_y=0$. The latter is possible due to rotational symmetry in the $xy$ plane. In particular, this shows that the material is birefringent.

\subsection{Reflection matrix in the chiral basis}

Since our goal is to calculate Casimir forces using Lifshitz's formula, we are interested in calculating the reflection matrix of a WSM that occupies the $z>0$ half space.
This means that the vacuum dispersion relation $\omega^2_{\rm vac}=k^2$ holds for $z<0$, whereas Eq.~\eqref{eq:disp} holds for $z>0$.

Since the system has translational symmetry in the $xy$ plane as well as on the time axis, $k_x$, $k_y$ and $\omega$ must be the same on both sides of the interface, and the only variable that can change is $k_z$. For an incoming wave with wave vector $\textbf{q}=(q_x,q_y,q_z)$ and frequency $\omega$, the outgoing wave inside the WSM will have the same frequency but a different wave vector $\textbf{k}=(q_x,q_y,k_z)$. Matching both dispersion relations, $\omega_{\rm vac}(\textbf{q}) = \omega_\pm(\textbf{k})$, there are two possible outgoing wave vectors $\textbf{k}^\pm$ for a given incoming wave vector $\textbf{q}$,
\begin{equation}
\label{eq:kz}
(k_z^\pm)^2=q_z \left(q_z \pm \sigma_{xy}\right) \, .
\end{equation}
Note that both of the two equations $\omega_{\rm vac}(\textbf{q}) = \omega_\pm(\textbf{k})$  have the same two solutions for $(k_z^\pm)^2$. Nevertheless, for incident wave vectors with $q_z < \sigma_{xy}$, there is only one polarization propagating inside the material, since the other polarization results in an evanescent wave \cite{grushin2012consequences}.

Thus, an incident wave with frequency $\omega$ and wave vector $\textbf{q}$ can be transmitted with two different wave vectors inside the WSL. Correspondingly, each of these wave vectors will be associated with a particular polarization $\textbf{e}_\pm$ of the transmitted field, which can be obtained from Eq.~\eqref{eq:polarizations},
%% are, for the displacement field,:
\begin{align} \label{eq:es}
\textbf{e}^\pm(\textbf{q}) =  q_z^2  \hat{x}
\mp i \omega q_z \hat{y}
 -  q_x k_z^\pm  \hat{z} \, .
\end{align}

In addition to energy conservation and momentum conservation perpendicular to the interface, to obtain the reflection matrix we demand that the electromagnetic field fulfils specific boundary conditions, namely, the continuity of the components $\textbf{E}_\parallel$ and $\textbf{B}_\parallel$ parallel to the interface.

Let us consider an incoming wave with ${\textbf{q}=(q_x,0,q_z)}$ that is either reflected at the interface to a new wave vector ${\textbf{q}_r=(q_x,0,-q_z)}$, or transmitted into the WSL with either of the wave vectors $\textbf{k}_\pm=(q_x,0,k_z^\pm)$. To obtain the reflection matrix, we need to match the incoming, reflected and transmitted electric and magnetic fields.

Inside the WSL, there is only one allowed polarization vector $\textbf{e}^\pm$ for a given wave vector $\textbf{k}^\pm$. Outside the WSM, however, there are two possible polarizations for a given wave vector, and different bases can be chosen that would each be correct solutions of Maxwell's equations. The most usual choice is the transverse basis $\lbrace \hat{e}_1,\hat{e}_2\rbrace$ that was
introduced in the end of last Section. However, this is not necessarily the best choice for every system: since ours has rotational symmetry about the $z$ axis, the most convenient choice of basis for the electromagnetic field is the chiral or circularly polarized basis,
\begin{align}
\hat{e}_{R,L} =& \frac{1}{2} \left( \hat{e}_1 \pm i \hat{e}_2 \right) = \frac{1}{\sqrt{2}} \left( \frac{q_z}{q} \hat{x} \pm i \hat{y} - \frac{q_x}{q} \hat{z} \right) \, , \label{eq:eRL} \\
\hat{e}_{R,L}^\prime =&  \frac{1}{2} \left( \hat{e}_1^\prime \pm i \hat{e}_2^\prime \right) = \frac{1}{\sqrt{2}} \left( -\frac{q_z}{q} \hat{x} \pm i \hat{y} - \frac{q_x}{q} \hat{z} \right) \label{eq:epRL} \, ,
\end{align}
where $q = |\textbf{q}|$ and $\hat{e}_{R,L}$ are the polarization vectors for circular waves with positive or negative helicity, propagating from left to right (from plate $A$ to plate $B$, see Fig. \ref{fig:figure}). On the other hand, $\hat{e}_{R,L}^\prime$ correspond to the case
where waves propagate from right to left (plate $B$ to plate $A$), with $\hat{e}_1^\prime= \hat{y} \times \hat{q}_r$ and $\hat{e}^\prime_2 = \hat{y}$.
It is easy to see that both $\lbrace \hat{e}_1, \hat{e}_2, \hat{q}_0 \rbrace$ and $\lbrace \hat{e}_1^\prime, \hat{e}_2^\prime, \hat{q}_r \rbrace$ form a right-handed set of vectors.

To understand the difference between $\hat{e}_{R,L}$ and $\hat{e}_{R,L}^\prime$ and their physical meaning, let us look at the case of normal incidence where $q_x=0$. In this case, when the incoming wave propagates along the $+\hat{z}$ direction,
we have $\textbf{q}_0=q_z \hat{z}$ with $q_z>0$ and $\hat{e}_{R,L} = ( \hat{x} \pm i \hat{y})/\sqrt{2}$.
We consider an incoming right-circularly polarized
field $\textbf{E}_0(\textbf{r}, t)=E_0  e^{i \textbf{q}_0 \cdot \textbf{r} - i \omega t} \hat{e}_R$. If this wave is reflected, its propagation direction will be reversed,
so the wave vector changes to $\textbf{q}_r = - q_z \hat{z}$.

As we mentioned, when the propagating direction is reversed, a right-handed set of vectors will be given by $\lbrace \hat{e}_1^\prime, \hat{e}_2^\prime, \hat{q}_r \rbrace$,
and the correct basis to describe the polarization of our system is the primed one \eqref{eq:epRL} that, for normal incidence, becomes
$\hat{e}_{R,L}^\prime =  (- \frac{q_z}{q} \hat{x} \pm i \hat{y})/{\sqrt{2}}$.
The reflected field can then be written as $\textbf{E}_r=E_0 e^{i \textbf{q}_r \cdot \textbf{r} - i \omega t} e^{i \pi}  \hat{e}_L^\prime$.
Thus we show that, for the normal incidence case,
a right-handed circularly polarized wave (positive helicity) incoming along the $+\hat{z}$ direction will become left-handed (negative helicity) and gain a $\pi$ phase when the direction of motion is reversed (as is the case, for instance, when reflected by a perfect mirror).

Thus we choose Eq.~\eqref{eq:eRL} as the basis for the incoming electric field $\textbf{E}_0$, Eq.~\eqref{eq:epRL} for the reflected field $\textbf{E}_{r}$, and Eq.~\eqref{eq:es} for the transmitted fields $\textbf{E}_\pm$. This allows us to write the total electric field at each side of the interface as
\begin{align}
&\textbf{E}_0=  E_0^R \hat{e}_R + E_0^L \hat{e}_L \nonumber \\
&\textbf{E}_r =  E_r^R \hat{e}_R^\prime + E_r^L \hat{e}_L^\prime \label{eq:Er} \\
&\textbf{E}_{\pm} =  E_{\pm} \left(  q_z^2 \hat{x} \mp i \omega q_z \hat{y} - q_x k_z^\pm \hat{z} \right)  \nonumber \, .
\end{align}

The magnetic field at each side of the interface can be readily written by taking $\textbf{B}_0 = \omega \textbf{q} \times \textbf{E}_0$, $\textbf{B}_r = \omega \textbf{q}_r \times \textbf{E}_r$ and ${\textbf{B}_\pm = \omega \textbf{k}_\pm \times \textbf{E}_\pm}$. Thus the boundary conditions can be used to obtain a linear system of four equations, one for the continuity of each field in the $x$ and $y$ directions. The resulting system of linear equations can be easily solved for the reflection coefficients $R_{ij}\equiv E_r^i/E_0^j$ with $i,j \in \{ R,L \}$. One finds that the reflection matrix for the WSM in the chiral basis has a simple off-diagonal form
\begin{equation}
\label{eq:Rrealchiral}
R(q_z) = \frac{1}{\sigma_{x y}} \left(
\begin{array}{c c}
0 & \sigma_{xy} + 2 k_z^- - 2 q_z \\
\sigma_{xy} - 2 k_z^+ + 2 q_z & 0
\end{array}
\right) \, .
\end{equation}
According to Eq.~\eqref{eq:Er}, this reflection matrix acts on an incoming vector in the basis $\lbrace\hat{e}_L,\hat{e}_R\rbrace$ and returns the components of the reflected vector in the basis $\lbrace\hat{e}_L^\prime,\hat{e}_R^\prime\rbrace$. This is the natural choice for the chiral basis associated with the two different propagation directions, as has been discussed above.

In order to obtain the correct Casimir force via Lifshitz's formula, it is necessary to obtain as well the mirrored reflection matrix $R'$ which accounts for the reflection of waves incoming towards a WSM occupying the $z<0$ ha  lf-space \cite{fialkovsky2018quest}. To do that it is necessary to consider an incoming wave with wave vector ${\textbf{q}_0^\prime=(q_x,0,-q_z)}$, a reflected wave with ${\textbf{q}_r^\prime=(q_x,0,q_z)}$, and transmitted fields with $\textbf{k}_\pm^\prime=(q_x,0,-k_z^\pm)$. Then, all the steps to obtain the reflection matrix can be reproduced, and the net effect of this change is that the roles of $k_z^+$ and $k_z^-$ are swapped, leading to the following mirrored reflection matrix, now relating an incoming vector in the primed chiral basis $\{ \hat{e}'_L, \hat{e}'_R\}$ with a reflected field written in the non-primed basis $\{ \hat{e}_L, \hat{e}_R \}$,
\begin{equation}
\label{eq:Rrealchiral2}
R'(q_z) = \frac{1}{\sigma_{x y}} \left(
\begin{array}{c c}
0 & \sigma_{xy} - 2 k_z^+ + 2 q_z  \\
  \sigma_{xy} + 2 k_z^- - 2 q_z& 0
\end{array}
\right) \, .
\end{equation}

\subsection{Wick rotation to imaginary frequencies}

In order to apply Lifshitz's formula, it is necessary to perform a Wick rotation to imaginary frequencies. This rotation has been performed in Ref.~\cite{wilson2015repulsive} but only under the assumption $\sigma_{xy}>0$. Allowing for the possibility of the cones being split in the opposite direction, i.e., $\textbf{b}= - b \hat{z}$, we will consider the case $\sigma_{xy}<0$ as well. This is relevant because for a given sample the sign of its Hall conductivity changes if it is mirrored in space \cite{burkov2014anomalousWeyl,nagaoisa2010anomalousHE}.

Defining $K_\pm = k^+_z \pm k^-_z$, we can write the off-diagonal elements of the reflection matrices as
\begin{align}
R_{LR}(q_z) &= \frac{1}{\sigma_{x y}} \left(  \sigma_{xy} -K_- + K_+ - 2 q_z \right)\, , \\
R_{RL}(q_z) &= \frac{1}{\sigma_{x y}} \left(  \sigma_{xy} - K_- - K_+ + 2 q_z \right) \, .
\end{align}
where
\begin{align}
    K_+ &= \sqrt{2 q_z \left( q_z + \sqrt{q_z^2 - \sigma_{xy}^2}\right)}\, ,  \notag \\
    K_- &= \text{sgn}({\sigma_{xy}}) \sqrt{2 q_z \left( q_z - \sqrt{q_z^2 - \sigma_{xy}^2}\right)}\, .
\end{align}
Therefore, $K_+$ is always positive, whereas $K_-$ changes sign with $\sigma_{xy}$. With these definitions we are able to rotate to imaginary frequencies,
defining $\omega = i \xi$, so that $q_z^2 = \omega^2 - q_x^2 - q_y^2 = -\xi^2 - q_x^2 - q_y^2$. This allows us to define $q_z = i p_z$, with $p_z \in \mathbb{R}$, fulfilling
$p_z^2=\xi^2 + q_x^2 + q_y^2 > 0 $. Then we obtain for $K_+$ and $K_-$,
\begin{align}
K_+  &= i \sqrt{2 p_z \left(\sqrt{p_z^2+\sigma_{xy}^2} + p_z \right)} \, , \\
K_- &= \text{sgn}(\sigma_{xy}) \sqrt{2 p_z \left(\sqrt{p_z^2+\sigma_{xy}^2} - p_z \right)} \, .
\end{align}
With this we can now write the reflection matrix in terms of the imaginary transversal momentum $p_z$,
so that the rotated reflection matrix in the chiral basis can be written as
\begin{equation}
\label{eq:Rimag}
R(i p_z)=
\left(
\begin{array}{c c}
0 & \mathcal{R}(p_z) \\
\mathcal{R}^*(p_z) & 0
\end{array}
\right) \,  ,
\end{equation}
with
\begin{align} \label{eq:func}
\mathcal{R}(p_z) &= 1 - g(p_z) + i h(p_z) \\
g(p_z) & = \frac{1}{|\sigma_{xy}|} \sqrt{2 p_z \left(\sqrt{p_z^2+\sigma_{xy}^2} - p_z \right)} \\
h(p_z) &= 	\frac{1}{\sigma_{xy}}  \sqrt{2 p_z \left(\sqrt{p_z^2+\sigma_{xy}^2} + p_z \right)} - 2 \frac{ p_z}{\sigma_{xy}}
\end{align}
Then, for the reflection matrix of the mirror-reflected system, we obtain
\begin{equation}
\label{eq:Rimag2}
R'(i p_z)=
\left(
\begin{array}{c c}
0 & \mathcal{R}^*(p_z) \\
\mathcal{R}(p_z) & 0
\end{array}
\right) \,  .
\end{equation}

\section{Chiral media}\label{sec:chiralmedia}

The other necessary element to describe our system is the chiral medium filling the gap between the two WSM. In a chiral medium, the eigenmodes are not TE-TM waves, but rather chiral states. Moreover, waves with different chiralities propagate in the medium with different velocities. The properties of the medium between the WSLs enters in the translation matrices in Eq.~\eqref{eq:Lif}. They are diagonal in the chiral basis.
Specifically, $U_{BA}$ takes a vector from the basis $\lbrace \hat{e}_L, \hat{e}_R\rbrace$ and gives a vector written in the primed basis $\lbrace \hat{e}_L^\prime, \hat{e}_R^\prime\rbrace$, while $U_{AB}$ does the opposite. They are given by \cite{jiang2019chiral}
\begin{equation}
U_{BA}=\left(
\begin{array}{c c}
e^{i k^+_L a} & 0 \\
0 & e^{i k^+_R a}
\end{array}
\right) , \, \,
U_{AB}=\left(
\begin{array}{c c}
e^{i k^-_L a} & 0 \\
0 & e^{i k^-_R a}
\end{array}
\right) \, ,
\end{equation}
where $k^\pm_R$ ($k^\pm_L$) are the $z$-component of the wavevectors of right (left) circularly polarized photons propagating in the $+$ (right to left) or $-$ (left to right) direction.
In vacuum, $k^\pm_{R,L}$ are simply equal to $q_z$ since all photons propagate with the same velocity,
but in chiral media they develop a shift in their propagation velocity that might depend on their chirality, their propagation direction and
the properties of the material involved.

From this we see that the chiral basis
was not only the natural choice to work in this type of WSM, but also in chiral materials. These are not two unrelated facts: in Ref. \cite{fukushima2019anomalous} it has been pointed out that WSMs with cones splitting in the direction of propagation are a particular type of chiral medium because the two polarizations inside the material propagate with different velocities. In the normal incidence case, when $q_x = 0$ and $\omega = q_z$, we can see that the eigenstates inside the material \eqref{eq:es} reduce indeed to the chiral basis.

Chiral media can be separated into two large groups depending on whether the propagation velocity depends only on the chirality of the photons or also on their propagating direction. The first class, which as we shall see below is not relevant for our work, are the \textit{optically active materials}. These materials preserve time-reversal symmetry, and thus photons with opposite chirality would propagate with different velocity independently of their direction of propagation: $k_R^\pm=\bar{k}_z + \delta k_z$, and $k_L^\pm=\bar{k}_z - \delta k_z$,
where $\bar{k}_z = i \sqrt{\omega^2 + k_\parallel^2} \equiv i p_z$
is the mean wave vector of photons and $\delta k_z$ is the difference of the wave vectors, which depends on the properties of the medium.
For optical active materials it can be taken as $\delta k_z = \alpha_0 \rho$, where $\alpha_0$ is the specific rotation and $\rho$ is the mass concentration of optically active molecules \cite{kauznman1940opticalrotation}.

The second class consists of the materials that display the \textit{Faraday effect}, an optical analogue of the Hall effect discovered by Faraday \cite{faraday1846magnetization},
in which the light passing through a medium in a magnetic field experiences a rotation of its polarization. In such materials, the optical rotation angle is given by ${\theta = \mathcal{V} B l}$,
where $\mathcal{V}$ is the Verdet constant which characterizes the material, $l$ is the distance travelled by light, and $B$ is the component of the magnetic field in the direction of propagation.
For Faraday materials, the photons display a difference in their propagating velocity that depends on their chirality
as well as on their direction of propagation, i.e., $k_R^\pm = k_L^\mp = \bar{k}_z \pm \delta k_z$, with $\delta k_z = \mathcal{V} B$. We will see shortly that Faraday materials have a interesting consequences for the Casimir force between WSMs.

\section{Casimir force between two Weyl semimetals in a chiral medium} \label{sec:casimirforce}

With all the elements we have presented so far, we are now ready to consider the Casimir force between two WSM, with Hall conductivities $\sigma_{xy}^{A,B}$, occupying the two half-spaces $z<0$ and $z>a$ respectively, as was shown in Fig.~\ref{fig:figure}. The gap between them is filled by a chiral medium, and we shall obtain the Casimir force by calculating the Casimir energy via Lifshitz's formula \eqref{eq:Lif}.
The reflection matrices for the two plates are
\begin{align}
R_A(i p_z) &=
\begin{pmatrix}
0 & \mathcal{R}_A^*(p_z) \\
\mathcal{R}_A(p_z) & 0
\end{pmatrix}
\,  , \notag \\
R_B(i p_z) &=
\begin{pmatrix}
0 & \mathcal{R}_B(p_z) \\
\mathcal{R}_B^*(p_z) & 0
\end{pmatrix}
\,  ,
\end{align}
where $\mathcal{R}_{A,B}(p_z)= \mathcal{R}(\sigma_{xy} = \sigma_{xy}^{A,B})$ and with $\mathcal{R}$ as derived in Eq.~\eqref{eq:func}. The off-diagonal form of the reflection matrices for WSM implies that the matrix involved in the calculation of the Casimir force results diagonal,
\begin{align}
\label{eq:M}
\mathbb{M} &\equiv  \mathbb{I} - R_B U_{BA} R_A U_{AB} \\ \nonumber
&=
 \left(
\begin{array}{c c}
1 - \mathcal{R}_B \mathcal{R}_A \, e^{i( k^+_R + k_L^-) a}   & 0 \\
0 & 1 -\mathcal{R}_B^* \mathcal{R}_A^* \, e^{i( k^+_L + k_R^-) a}
\end{array}
\right) \, .
\end{align}
This matrix can be interpreted in the following way: the first component contains a term $\mathcal{R}_B e^{i k_R^+ a} \mathcal{R}_A e^{i k_L^-}$, which accounts for a left-polarized photon travelling in the $-$ direction towards the $A$ plate, where it is reflected. The Fresnel coefficient for such reflection is $\mathcal{R}_A$, i.e., the photon will be reflected with probability $\mathcal{R}_A$ and transmitted with probability $1-\mathcal{R}_A$. Given the rotational symmetry of the system, its angular momentum cannot be changed under reflection. However, since its linear momentum changes and the direction of motion is reversed, its helicity changes. It thus becomes a
photon with right chirality, travelling in the $+$ direction towards the $B$ plate. There, it again gets reflected with the Fresnel coefficient $\mathcal{R}_B$, and the propagation starts over.

The other non-vanishing component contains a term  $\mathcal{R}^*_B e^{i k_L^+ a} \mathcal{R}^*_A e^{i k_R^-}$. Here, we have a right-polarized photon which undergoes the same situation as before, but
this time the Fresnel coefficients involved in the reflections are the complex conjugates $\mathcal{R}^*_{A,B}$. This means that a WSM acts as a regular imperfect mirror for chiral photons, but the reflection coefficient \textit{depends on the
chirality of the photons}. If the photons were polarized in the TE-TM basis, then the WSM would mix their polarizations.

As we saw in Sec.~\ref{sec:chiralmedia}, there are two different types of chiral media.
For optically active media, we have seen that the velocity of photons does not depend on their propagation direction. By looking at Eq.~\eqref{eq:M} it can be seen that, since $k_R^\pm = \bar{k}_z + \delta k_z$ and $k_L^\pm = \bar{k}_z - \delta k_z$, all the dependence on the characteristics of the material ($\delta k_z$) would vanish for the Casimir energy. For this reason, we will focus from now on Faraday materials.

To obtain the Casimir force, we need to calculate
\begin{align}
m(p_z) & \equiv \det \left(\mathbb{M}\right)  \\
&=  1 - 2 e^{-2 p_z a} \left[ C_R(p_z) \cos\left( 2 \delta k_z a \right)  \right. \nonumber \\
&  \left. - \, C_I (p_z) \sin\left( 2 \delta k_z a \right) \right] + W(p_z) e^{-4 p_z a} \nonumber
\, ,
\end{align}
where we have defined
\begin{align}
W(p_z) &=  |\mathcal{R}_A|^2 |\mathcal{R}_B|^2   \\
&= \left[\left(1-g_A\right)^2 + h_A^2 \right] \left[ \left(1-g_B\right)^2 + h_B^2 \right] \nonumber\\
C_R(p_z) &=  \text{Re} \left(\mathcal{R}_A \mathcal{R}_B \right)
= (1-g_A)(1-g_B) - h_A h_B \nonumber \\
C_I(p_z) &= - \text{Im} (\mathcal{R}_A \mathcal{R}_B) = h_A (1-g_B) + h_B (1-g_A) \nonumber
\end{align}
where $g_{A,B}(p_z)$ and $h_{A,B}(p_z)$ are defined in Eq.~\eqref{eq:func}. We shall remain in a limit where neither $\delta k_z$ nor $\sigma_{xy}^{A,B}$ depend on the frequency or wavelength of the
light. Hence, a change of variables to spherical coordinates can be performed, noting that $p_z^2= \xi^2 + q_x^2 + q_y^2$, so that we can write
$q_x= p_z \cos \phi \sin \theta$, $q_y = p_z \sin \phi \sin \theta$, and $\xi = p_z \cos \theta$, with $\phi \in [0,2\pi]$, $\theta \in [0, \pi/2]$ and $p_z \in [0,\infty]$.
Then the Casimir force $F_C= - \partial E_c / \partial a$ can be written as
\begin{equation}
F_c = - \frac{\hbar}{(2 \pi)^2} \int_0^\infty dp_z \, \frac{p_z^2}{m(p_z)} \frac{\partial m(p_z)}{\partial a} \, .
\end{equation}%
\begin{figure}[t]
	\includegraphics[width=\columnwidth]{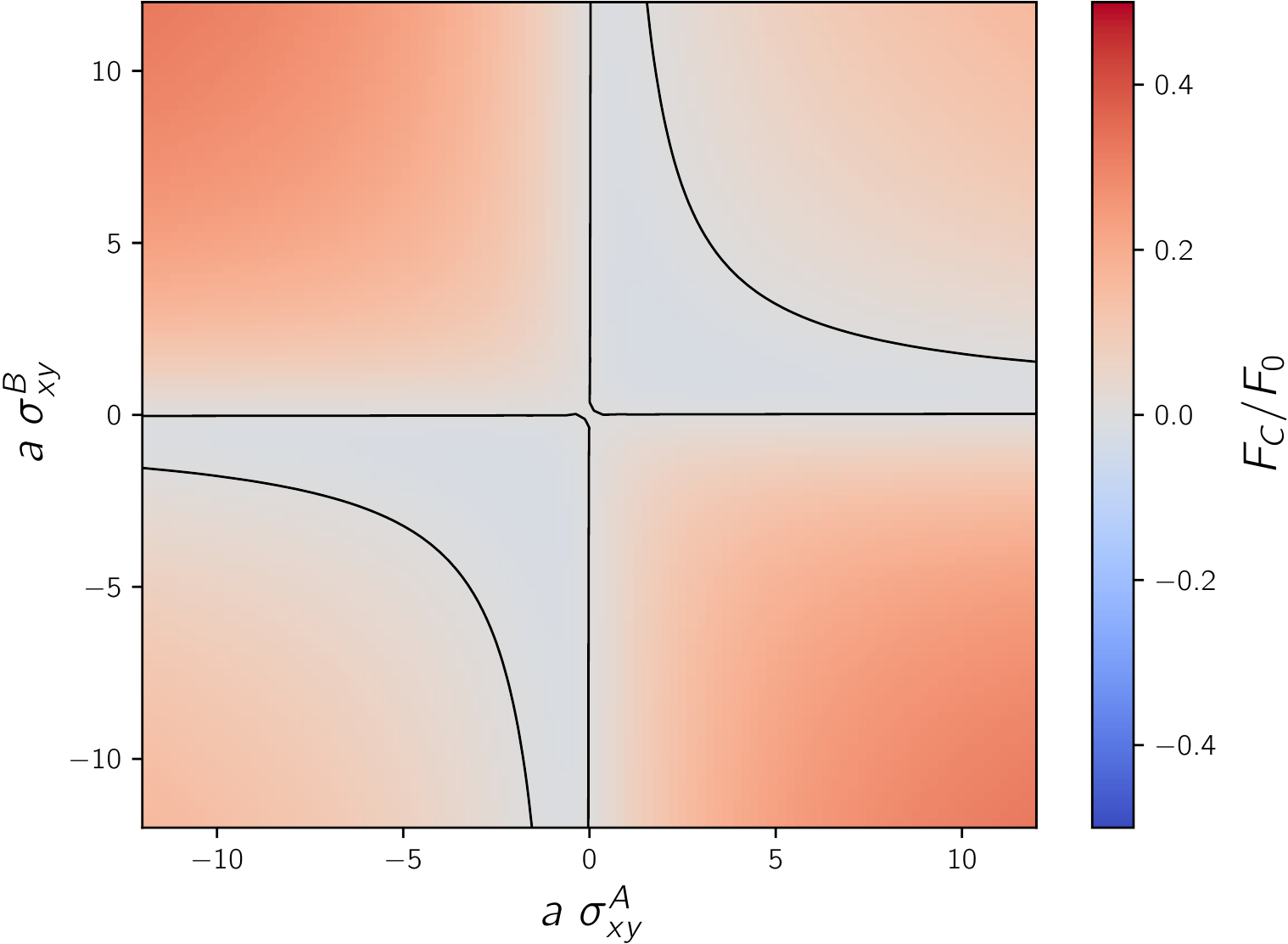}
	\caption{\label{fig:diffsigma} Casimir force for two WSMs with different Hall conductivities $\sigma_{xy}^A$ and $\sigma_{xy}^B$, with vacuum filling the gap between them. The Casimir force is shown as a function of the dimensionless conductivities $a \sigma^{A}_{xy}$ and $a \sigma^{B}_{xy}$, where $a$ is the distance between the plates. The black solid lines are the curves in which the force vanishes, and the light blue areas are the zones in which the Casimir force is repulsive.}
\end{figure}%
\begin{figure}[t]
	\includegraphics[width=\columnwidth]{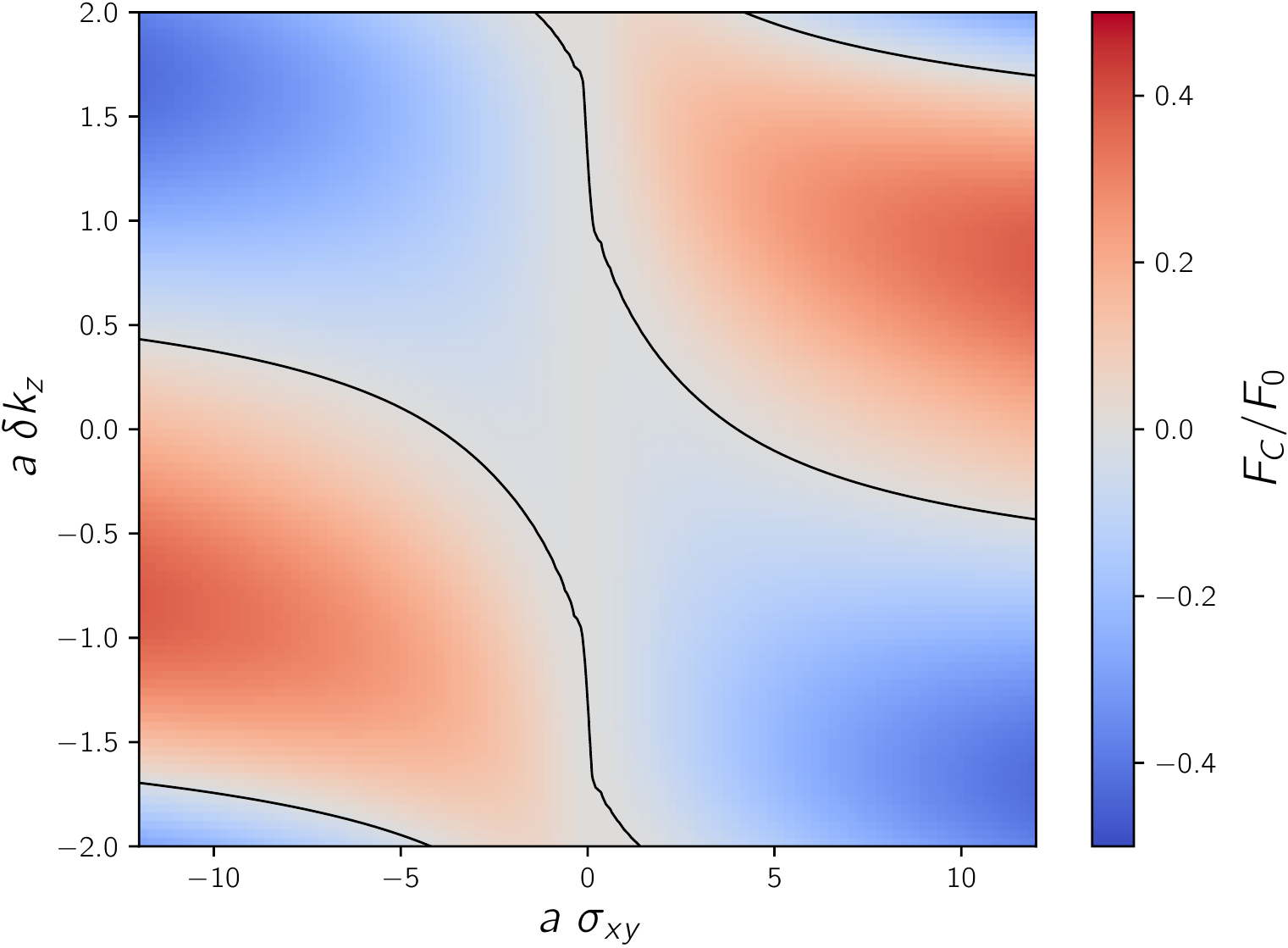}
	\caption{\label{fig:deltakz} Casimir force for two WSM with identical Hall conductivities $\sigma_{xy}^A = \sigma_{xy}^B = \sigma_{xy}$ when the gap in between them is filling by Faraday material.
		The force is shown as a function of $\sigma_{xy} a$ and $\delta k_z a$, where  $a$ is the distance between the plates and
		$\delta k_z$ is the difference between the propagating velocities of photons with different chirality, which is a characteristic of the chiral medium. The solid black lines correspond to a vanishing force, marking the transition from regions with attraction (in red) an repulsion (in blue).}
\end{figure}%
Now we can introduce the dimensionless variables ${u \equiv p_z a}$, as well as $\delta k_z a$ and $a \sigma_{xy}^{A,B} $ (the latter is dimensionless because we are using Gaussian units and taking $c=1$), so that the Casimir force results in
\begin{equation}
\frac{F_C}{F_0}= \frac{240}{\pi^4} \int_0^\infty du \, u^2 e^{- 2 u} \frac{n(u)}{m(u)} \, ,
\end{equation}
where $F_0 = - \hbar \pi^2 / 240 a^4$ is the attractive Casimir force between two perfectly conducting plates, and
\begin{align}
n(u) &= \,C_R(u) \left[ u \cos(2 a \delta k_z) + a \delta k_z \sin(2 a \delta k_z) \right] \\
&- C_I(u) \left[ u \sin(2 a \delta k_z) - a \delta k_z \cos(2 a \delta k_z) \right]\nonumber \\
&- u W(u) e^{-2 u} \nonumber  \,,
\end{align}
and the functions $m(u)$, $C_R(u)$, $C_I(u)$, $W(u)$, $r_{A,B}(u)$ and $p_{A,B}(u)$ are obtained by replacing $p_z \rightarrow u$, $\sigma_{xy}^{A,B} \rightarrow a \sigma_{xy}^{A,B}$, and $\delta k_z \rightarrow a \delta k_z$ in their original definitions.

At this point it is easy to check some basic limits: the limit $a \sigma_{xy}^{i} \rightarrow 0$ for $i \in \{A,B\}$ results in $r_i \rightarrow 1$ and $p_i \rightarrow 0$, so that the force vanishes. This was expected because in that limit the WSM effectively behaves like vacuum as can be seen from its dielectric constant, Eq.~\eqref{eq:eps}.

On the other hand, when $a \sigma^{i}_{xy} \rightarrow \infty$, one finds $r_i,p_i \rightarrow 0$, so the reflection matrix in the chiral basis has ones in the off-diagonal elements, which would correspond to a perfectly conducting plate. And indeed, we can check that in this limit of $F_c/F_0 \rightarrow 1$.

\begin{figure*}[t]
	\includegraphics[width=18cm]{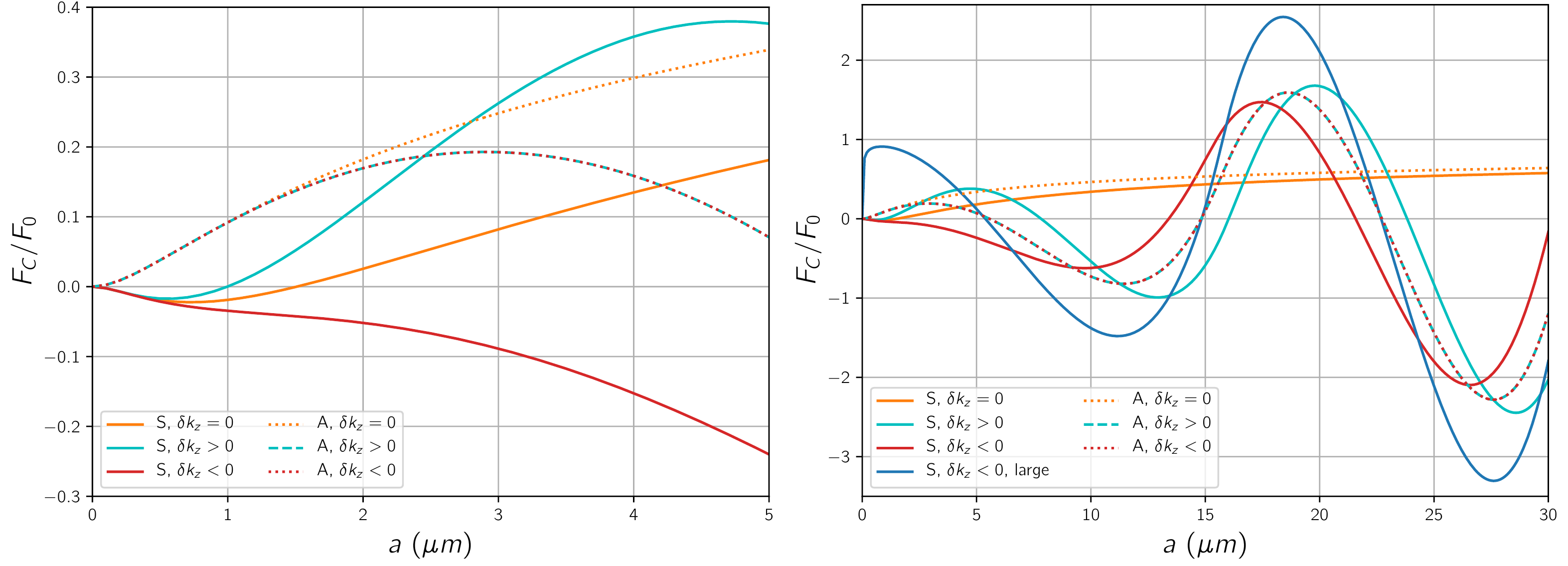}
	\caption{\label{fig:distance} Casimir force as a function of the distance, for long and short distances. The $\delta k_z = 0$ case corresponds to vacuum filling the gap between the two WSMs. The nonvanishing $\delta k_z$ are set to be $\delta k_z= \pm 2 \times 10^5 m^{-1}$, and the values of the Hall conductivities of the WSMs are such that $\sigma_{xy}^A/c=2.65 \times 10^{6} m^{-1}$.
		%The dashed gray line marks the transition from attractive to repulsive.
		The curves marked with a $S$ are the ones corresponding to a symmetric configuration where $\sigma_{xy}^A = \sigma_{xy}^B$, while $A$ stands for the antisymmetric configuration with $\sigma_{xy}^A = - \sigma_{xy}^B$.
		The line marked ``\textit{large}'', shown in the plot for large distances, corresponds to a value $\sigma_{xy}^A/c=2650 \times 10^{6} m^{-1}$ and illustrates the perfect conductor limit.}
\end{figure*}

We will begin by considering the case in which the gap between the two WSM is filled by vacuum, as considered in Ref.~\cite{wilson2015repulsive}, but allowing for the Hall conductivities of the plates to be different from one another.
We show these results in Fig.~\ref{fig:diffsigma}. We see that a necessary condition for repulsion to be possible, in absence of a chiral medium, is that $\text{sgn}(\sigma_A)=\text{sgn}(\sigma_B)$. Moreover, one sees that for fixed distance, as long as one of the Hall conductivities is small enough, a repulsive force can still be obtained for large values of the other Hall conductivity.
However, it is worth pointing out that even though repulsion can be achieved in vacuum, its magnitude is still only about $5\%$ of the magnitude of the Casimir
force between two perfectly conducting plates for the parameters considered. Nevertheless, a significant reduction of the attractive force can be achieved.

Another interesting effect can be seen already in Fig.~\ref{fig:diffsigma}. A change from $\sigma_{xy}$ to $- \sigma_{xy}$ can be obtained trivially by simply flipping the sample as this results in a reversal of the separation between Weyl cones from $\textbf{b}=b \hat{z}$ to $\textbf{b}=-b \hat{z}$. This means that at a fixed distance and for a given pair of samples, flipping one of the samples would result in a change from attraction to repulsion and vice versa. Being capable of achieving both repulsion and attraction without the need of changing the distance, applying external fields or doping, makes WSL a very versatile testbed for studying the Casimir repulsion.

Next, we turn to the case of a chiral medium filling the gap between the WSMs. In Fig.~\ref{fig:deltakz} we consider the case in which both WSMs have the same Hall conductivity $\sigma_{xy}^A = \sigma_{xy}^B = \sigma_{xy}$, and show the Casimir
force as a function of the dimensionless variables $\sigma_{xy} a$ and $\delta k_z a$, where $\delta k_z$ is the shift in propagating velocities of photons with different chirality. The dashed black line separates
the region in which the force is attractive from the region where repulsion is achieved. For a fixed distance between the plates, we can see that the presence of the chiral medium can lead to stronger repulsive forces, reaching about $50 \%$ of the magnitude of the Casimir force between two perfectly conducting plates $F_0$ for the parameters considered.

On the other hand, we know that the sign of $\delta k_z$ indicates the chirality that propagates faster than the other within the Faraday material.
That is, for positive $\delta k_z$, a right-handed wave moving in the $+\hat{z}$ direction will have  a larger propagating velocity than a
left-circularly polarized wave moving in the same direction. If $\delta k_z$ is negative, such relation will be inverted. From Fig.~\ref{fig:deltakz} we can see that the repulsion is enhanced in the regions in which the sign of $\delta k_z$ is opposite to the sign of $\sigma_{xy}$. The sign of $\delta k_z$
can be changed by flipping the orientation of the external magnetic field.

We can gain a deeper insight into the effect of both the WSM and the Faraday material by looking at the dependence of the force on the distance between the plates. To do so we rely on the available experimental data: for the WSM, we will consider a Hall conductivity of $\sigma_{xy}/c=2.65 \times 10^6 \text{m}^{-1}$ in Gaussian units, which corresponds to the measured $\sigma_{xy}=870 \Omega^{-1} \text{cm}^{-1}$ \cite{Belopolski1278}. Regarding the Faraday material, we will follow Ref.~\cite{jiang2019chiral} and consider the large Verdet constant $\mathcal{V}=5 \times 10^4 \text{rad} \text{m}^{-1} \text{T}^{-1}$, and an external applied magnetic field of $B=4 \text{T}$, which results in a phase difference $\delta k_z = \mathcal{V} B \approx 2 \times 10^5 \text{rad} \, \text{m}^{-1}$ between the left- and right-circularly polarized photons.

\begin{figure}[t]
	\includegraphics[width=\columnwidth]{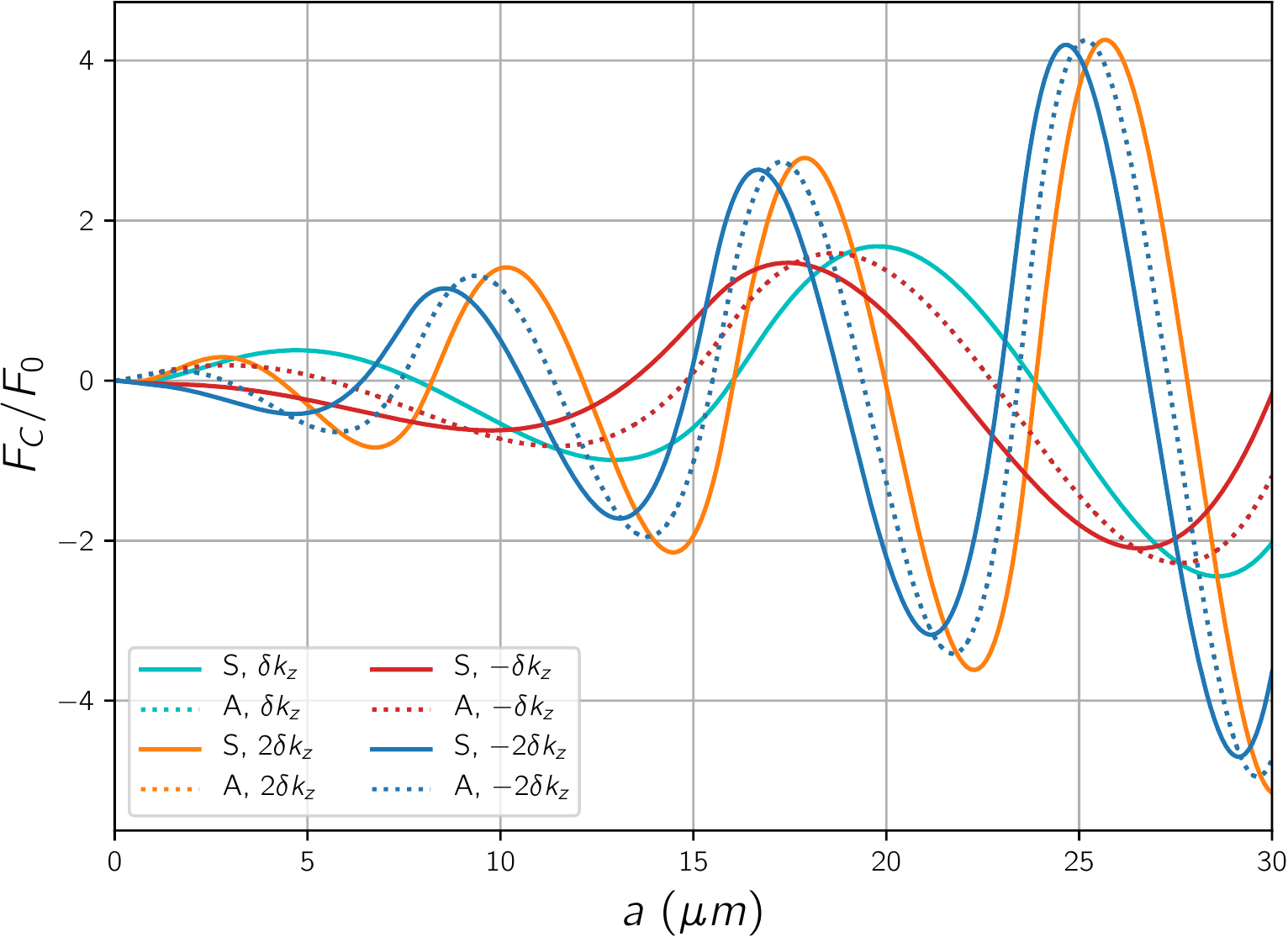}
	\caption{\label{fig:distancedkz} Effect of the sign and amplitude of $\delta k_z$ on the Casimir force for long distances. The  values of the Hall conductivities of the WSMs are such that $\sigma_{xy}^A/c=2.65 \times 10^{6} m^{-1}$, and $\delta k_z= \pm 2 \times 10^5 m^{-1}$.
		%The dashed gray line marks the transition from attractive to repulsive.
	The curves marked with a $S$ are the ones corresponding to a symmetric configuration where $\sigma_{xy}^A = \sigma_{xy}^B$, while $A$ stands for the antisymmetric configuration with $\sigma_{xy}^A = - \sigma_{xy}^B$.}
\end{figure}

In Fig.~\ref{fig:distance} we show the behavior of the Casimir force for both short ($0-5 \mu$m) and long ($0-30 \mu$m) distances.
For short distances, we can see what we already observed from Fig. \ref{fig:deltakz}: that there is no Casimir repulsion between WSMs whose
conductivities have different signs, regardless of the presence of a Faraday medium in between them (light blue and orange curves).
A flip in the sign of the phase velocity difference $\delta k_z$ would not affect the force in this case (as will be shown in Fig. \ref{fig:distancedkz}),
since both systems would be equivalent. In contrast, when the conductivities of both WSMs have the same sign, then the introduction of a Faraday material with a positive
Verdet constant and the application of an external magnetic field in the opposite direction to the splitting of the Weyl cones would lead to an
enhancement of the repulsion (light purple curve). A system with this characteristics would be able to host repulsive forces even at small distances
(in contrast with the perfectly conducting case \cite{jiang2019chiral}), but the range of distances for which repulsion is achieved is much larger than in the absence of the Faraday medium.
The magnitude of the repulsive force, for small distances, can be as well much larger than the repulsive force obtained in vacuum.

When we have a look at larger distances, the first thing we notice is that, in the absence of chiral medium (light blue curves),
the only repulsive region exists at small distances, and there is only one
trapping distance where the force vanishes. The presence of the chiral medium is what creates the oscillations already found in
Ref.~\cite{jiang2019chiral} for the force between two perfectly conducting plates.
We show this limit as well, using a very large value for the Hall conductivity, as a dark
blue line. We can see that the main advantage of having a finite Hall conductivity via WSMs is found at small distances,
for which repulsion can only be achieved in this scenario. At larger distances, the finite Hall conductivity changes slightly the position of the zeros,
allowing for more control on the position of the trapping distances.

Lastly, in Fig.~\ref{fig:distancedkz}, we show the behavior of the force for different values and signs of $\delta k_z$. We find that the frequency of the oscillations grows with $\delta k_z$, and that for long distances, the sign change only moves slightly the position of the trapping distances.

\section{Conclusions}\label{sec:conc}

In this work we have studied the Casimir interaction between two semi-infinite Weyl semimetals separated by a distance $a$, and whose Weyl cones are split in the
same direction as the gap between the two materials. This fact renders both plates rotationally symmetric, and we found that this implies
that they show a particular response in presence of circularly polarized light:
each photon sees the WSM as an imperfect isotropic mirror, but with a different reflection coefficient depending on its chirality.

Even if the gap is filled with vacuum, WSMs exhibit interesting features regarding Casimir interactions, as has been shown in the literature \cite{wilson2015repulsive,rodriguez2019signatures}.
We found that a flipping of one sample in real space may switch the system from repulsive to attractive and vice-versa,
which makes WSMs very versatile materials from the point of view of investigating the Casimir force, given the fact that the same sample may exhibit either repulsion
or attraction depending only on its orientation with respect to another sample.

We then considered the case in which the gap between the two material is filled with a chiral medium, either a Faraday material or an optically active medium.
We found that, in the latter case, the medium has no effect on the Casimir force. In contrast, the presence of a material with Faraday rotation between the two WSMs presents several advantages with a view towards Casimir repulsion.

The addition of a Faraday medium in the gap enhances the repulsive force at short distances,
as long as the magnetic field that generates the difference between the propagating velocities in the Faraday material points in the opposite direction
as the separation between the Weyl cones in momentum space. The effect of the external magnetic field on the WSM can be neglected.
It is worth point out, however that for large separation between the plates
using highly conducting metals would lead to stronger repulsion than using WSMs \cite{jiang2019chiral}.

The system we presented here allows for the manipulation of various parameters, some intrinsic to the materials employed (the absolute value of the Hall
conductivity $|\sigma_{xy}|$ of the WSM, the Verdet constant of the Faraday material $\mathcal{V}$), but also some external ones like the orientation of the
plates which determines the sign of their conductivity and the external magnetic field. Suitable combinations of these parameters can be used to place
the trapping distance, in which no force acts on the plates, at any desired distance between them.

\section{Acknowledgements}
A.Z. is supported by the Academy of Finland Grant No. 308339. M.B.F. and T.L.S. acknowledge financial support from the National Research Fund Luxembourg under Grants CORE~11352881 and ATTRACT~7556175.

%%%%%%%%%%%%%%%%%%%%%%%%%%%%%%%%%%%%%%
\bibliography{casimirTI}

\end{document}